\documentclass[journal=nalefd, manuscript=letter,
email=true]{achemso}
\usepackage[english]{babel}
\usepackage{amsmath}
\usepackage[colorlinks=true, pdfstartview=FitV, linkcolor=blue, citecolor=blue, urlcolor=blue]{hyperref}

\newcommand{\ket}[1]{\left|{#1}\right\rangle}
\newcommand{\bra}[1]{\left\langle{#1}\right|}

\author{M. Montinaro}

\author{G. W\"ust}

\author{M. Munsch}

\affiliation{Department of Physics, University of Basel, Klingelbergstrasse 82, 4056 Basel, Switzerland}

\author{Y. Fontana}

\author{E. Russo-Averchi}

\author{M. Heiss}

\affiliation{Laboratoire des Mat\'eriaux Semiconducteurs, \'Ecole Polytechnique F\'ed\'erale de Lausanne, 1015 Lausanne, Switzerland}

\author{A. Fontcuberta i Morral}

\affiliation{Laboratoire des Mat\'eriaux Semiconducteurs, \'Ecole Polytechnique F\'ed\'erale de Lausanne, 1015 Lausanne, Switzerland}

\author{R. J. Warburton}

\author{M. Poggio}

\email{martino.poggio@unibas.ch}

\affiliation{Department of Physics, University of Basel, Klingelbergstrasse 82, 4056 Basel, Switzerland}

\title{Quantum dot opto-mechanics in a fully self-assembled nanowire}

\date{\today}

\begin{document}


\begin{abstract} 
  We show that fully self-assembled optically-active quantum dots (QDs)
  embedded in MBE-grown GaAs/AlGaAs core-shell nanowires (NWs)
  are coupled to the NW mechanical motion. Oscillations of the NW
  modulate the QD emission energy in a broad range exceeding 14~meV.
  Furthermore, this opto-mechanical interaction enables the dynamical
	tuning of two neighboring QDs into resonance, possibly allowing for
  emitter-emitter coupling. Both the QDs and the coupling mechanism
	-- material strain --
  are intrinsic to the NW structure and do not depend on any
  functionalization or external field.  Such systems open up the
  prospect of using QDs to probe and control the mechanical state of a
  NW, or conversely of making a quantum non-demolition readout of a QD
  state through a position measurement.
	
	\textbf{Keywords:} \textit{hybrid system, quantum dot, nanowire, self-assembly, strain, opto-mechanics.}
\end{abstract}





Experiments on micro- and nanomechanical oscillators are now addressing
what were once purely theoretical questions: the initialization,
control, and read-out of the quantum state of a mechanical oscillator.
Researchers are able both to initialize the fundamental vibrational
mode of a mechanical resonator into its ground state\cite{Teufel, Chan}
and even to produce non-classical coherent states of motion\cite{OConnell}.
The prospects are bright for exploiting these achievements to produce
mechanical sensors whose sensitivity is limited only by quantum effects
or to use a mechanical state to encode quantum information. The ability
to initialize and observe the quantization of mechanical motion is
particularly noteworthy not only from a fundamental point of view, but
also because mechanical oscillators are excellent transducers. By
functionalizing a resonator with an electrode, magnet, or mirror,
mechanical motion can be transformed into the modulation of electric,
magnetic, or optical fields\cite{Treutlein}. The ease of this process
has inspired proposals to use mechanical resonators as quantum
transducers, mediating interactions between different quantum
systems\cite{Rabl, Kolkowitz, McGee, Palomaki}. Furthermore, such
couplings have now been demonstrated in a variety of quantum systems
including optical\cite{Anetsberger} and microwave\cite{Bochmann}
cavities, superconducting devices\cite{Armour}, laser-cooled
atoms\cite{Camerer}, quantum dots\cite{Yeo} and nitrogen vacancy
centers in diamond\cite{Arcizet, Teissier, Jayich}. In most cases,
however, the functionalization of the mechanical oscillator with a
coupling element competes with the requirement of a small resonator
mass, required for achieving a high coupling strength\cite{Treutlein}.
Moreover, the functionalization process often adds additional paths of
dissipation and decoherence, reducing the lifetime of the coupled
quantum system, or ``hybrid'' system.

In this letter, we report on the coupling of a nanomechanical
oscillator with controllable quantum states, in which both the coupling
interaction and the quantum states themselves are intrinsic to the
oscillator's structure.  Not only is the strength of this coupling
unusually strong, but its ``built-in'' nature produces a hybrid system
whose inherent coherence is unspoiled by external functionalization
and whose fabrication is simpler than top-down techniques.  The
specific nanoresonator that we study is a bottom-up GaAs/AlGaAs
core-shell nanowire (NW) containing optically-active self-assembled
quantum dots (QDs)\cite{Heiss}.  These QDs have been shown to emit
narrow optical linewidth (down to 29~$\mu$eV) single photons with high
brightness (count rates in the MHz range) \cite{Heiss}. Here we show
that their energy levels are coupled to the mechanical vibrations of
the NW through intrinsic material strain.  We demonstrate that
mechanical motion allows a reversible tuning of the QD optical
frequency with no measurable influence on its photoluminescence
intensity.

\begin{figure}[tbhp]
	\includegraphics[width=6.5 in]{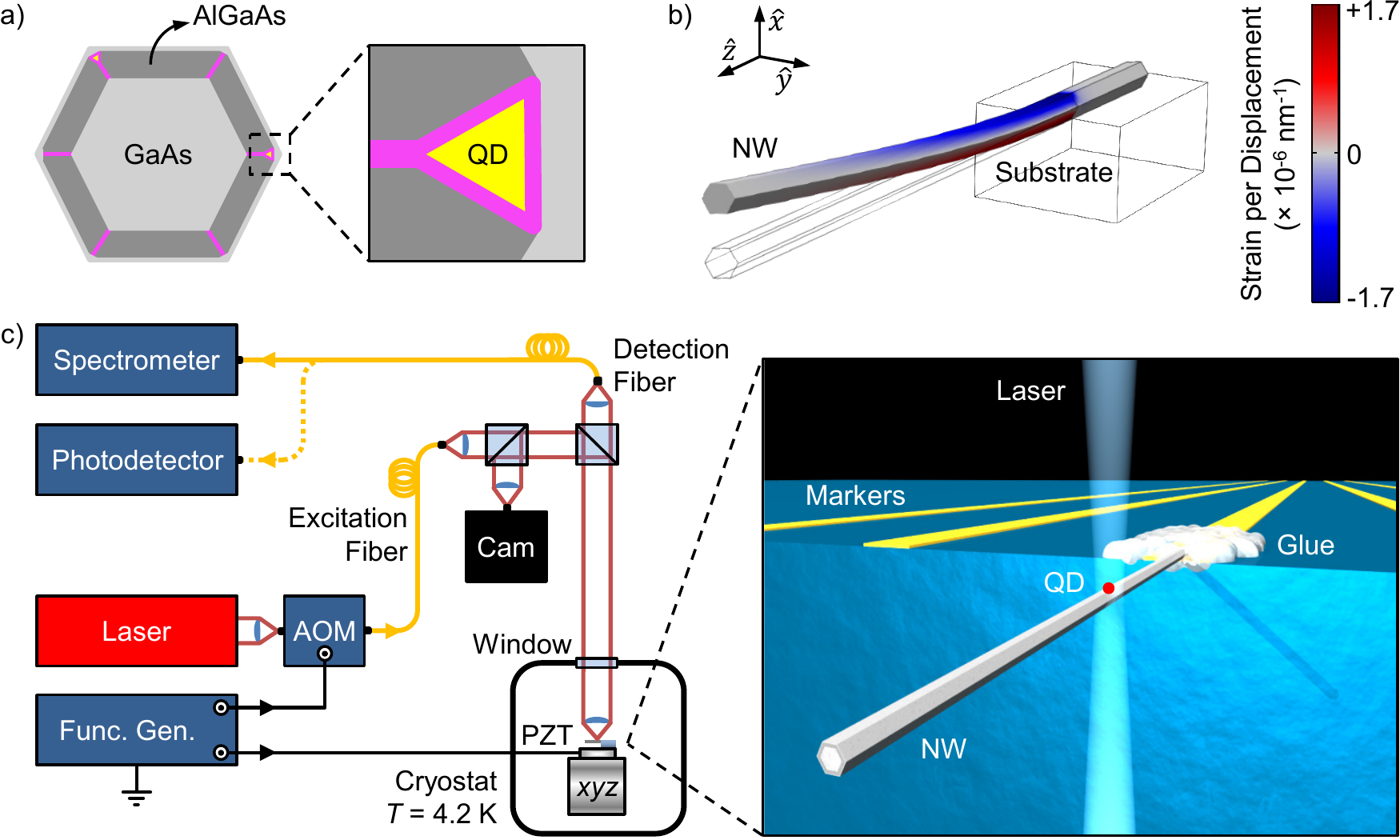}
	\caption{\label{f:setup} Experimental setup. (a)~Sketch of the
          NW cross-section, showing the composition of the core/shell
          structure and the close proximity of the QDs to the surface
          of the NW. The purple regions are rich in Al content and
          surround an Al-poor region (yellow), defining a QD. (b)~Finite
          element model of the displacement-dependent strain in the
          NW. The color scale is proportional to the
          $\varepsilon_{zz}$ component of the strain tensor
          $\varepsilon$, computed for the prominent flexural vibration
          along $\hat{x}$. (c)~Schematic diagram of the experimental
          setup.}
\end{figure}

Our quantum-dot-in-nanowire structures are fully self-assembled by
molecular beam epitaxy (MBE). As shown in Fig.~\ref{f:setup}(a), the
QDs form at the apex of the GaAs/AlGaAs interface, in Al-poor regions
embedded in the Al-rich corners of the NW hexagonal cross-section.  By
controlling the overall diameter of core and shell during growth, it
is possible to position the QDs within a few nanometers of the NW
surface. This proximity to the surface allows for the optimal coupling
of the QDs to the strain in the NW (Fig.~\ref{f:setup}(b)).  Despite
their position near the surface, these QDs retain their high optical
quality, making them ideal for sensing applications. The NWs studied
here have a predominantly Zinc-Blende crystalline structure and
display a regular hexagonal cross-section. The synthesis starts with a
290-nm thick NW core, grown along $[1\,\bar{1}\,1]$ on a Si substrate
by the Ga-assisted method detailed in Uccelli \textit{et al.} and
Russo-Averchi \textit{et al.} \cite{uccelli, Russo}.  Once the NWs are
about $25\,\mu$m long, the axial growth is stopped by temporarily
blocking the Ga flux and reducing the substrate temperature from 630
down to $465\,^{\circ}$C. Then a 50-nm thick Al$_{0.51}$Ga$_{0.49}$As
shell capped by a 5-nm GaAs layer is grown as detailed in Heigoldt
\textit{et al.} \cite{heigoldt}.


In order to study the opto-mechanical coupling, individual NWs are
detached from their growth substrate with micro-manipulators and
glued (using an ultra-violet curing adhesive) in a cantilever
configuration to the edge of a Si chip, which has been patterned with
lithographically defined alignment markers (Ti/Au, 10/30~nm thick).
The suspended length of the NWs typically amounts to 20~$\mu$m. The
chip is then rigidly fixed to a piezoelectric transducer (PZT),
which is used to excite mechanical oscillations of the NW, as shown
in Fig.~\ref{f:setup}(c). The chip and PZT are mounted to a
three-dimensional positioning stage which has nanometer precision and
stability (Attocube AG), in a low-pressure $^4$He chamber
($p = 0.35$~mbar) at the bottom of a $^4$He cryostat ($T = 4.2$~K).
The positioning stage allows precise alignment of individual QDs
within each NW with the 400-nm collection spot of a confocal
optical microscope\cite{Hoegele} with high numerical aperture
(NA $= 0.82$). As shown schematically in Fig.~\ref{f:setup}(c), the
microscope consists of a low-power, non-resonant HeNe excitation
laser at 632.8~nm, a camera for imaging the sample, and a
high-resolution spectrometer for analyzing the emitted
photoluminescence (PL). The mechanical oscillation of each NW is
detected via laser interferometry\cite{Bruland}. $80\,\mu$W of laser
light from a wavelength-tunable, highly coherent 780-nm laser diode are focused 
onto the NW free end and the reflected light is collected by a fast
photodetector. A low-finesse Fabry-P\'erot cavity, with a length of
$118 \pm 5$~cm, forms between the NW and a low-reflective window at
the entrance of the $^4$He chamber, as confirmed by a measurement
of its free spectral range. Measurements of the NW displacement by
the interferometer are calibrated by an accurate determination of
the laser wavelength (for more details, see supporting information).

\begin{figure}[tbhp]
	\includegraphics[width=3.33 in]{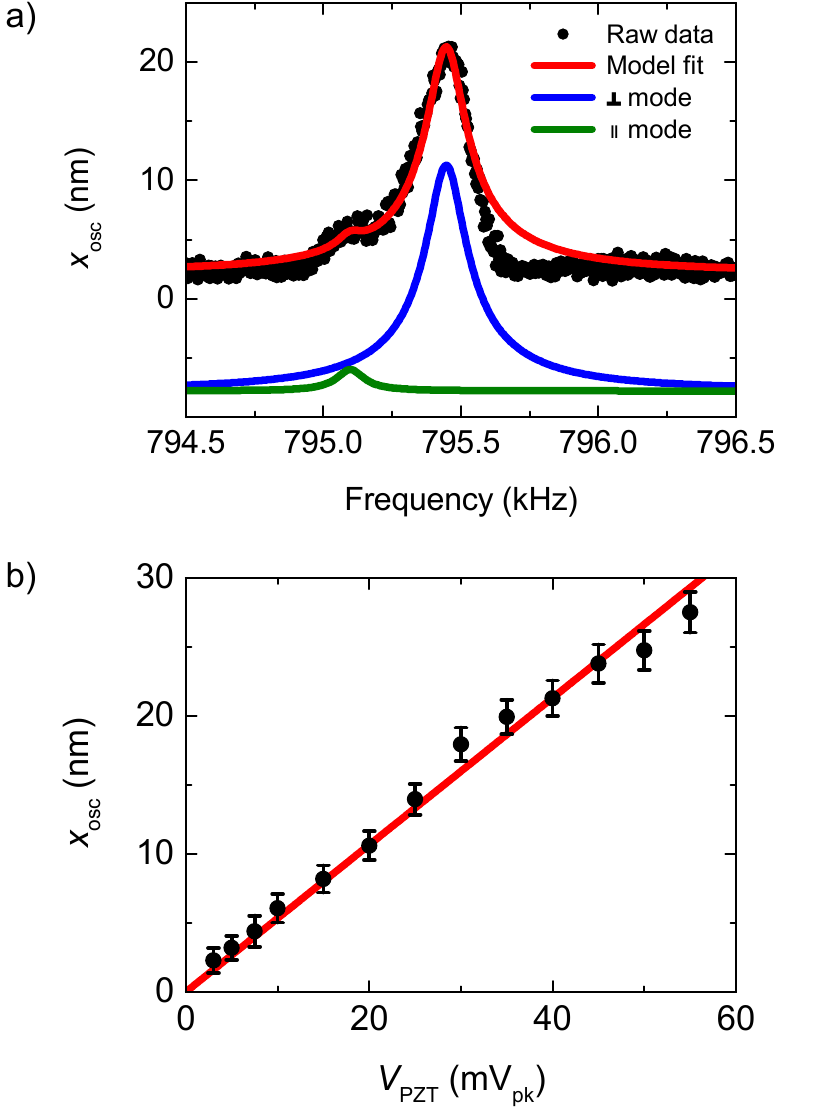}
	\caption{\label{f:interf} NW mechanical characterization.
	(a)~Spectrum of the NW free-end oscillation amplitude
	$x_{\text{osc}}$ corresponding to its lowest order flexural
	vibrations at $T = 4.2$~K, driven by the PZT at
	$V_{\text{PZT}} = 40$~mV$_{\text{pk}}$. The red line is a model
	fit (see main text), highlighting two resonances split by 350~Hz,
	corresponding to polarized, non-degenerate vibrational modes. The
	green curve represents the vibration parallel to the NW substrate,
	while the blue curve represents the perpendicular one (both are
	shifted for clarity). The mechanical quality factors of the two
	modes, extracted from the fit, are $Q_{\parallel} = 7600$ and
	$Q_{\perp} = 5800$. (b)~NW free-end oscillation amplitude
	$x_{\text{osc}}$ as a function of the amplitude of the PZT
	excitation voltage $V_{\text{PZT}}$. 
	The error bars correspond to the peak-to-peak amplitude of the
	interferometric noise. The red line is a linear fit, from
	which we extract the conversion factor
	$\partial x / \partial V_{\text{PZT}} = 0.53 \pm 0.01$~nm/mV.}
\end{figure}

Using the PZT, we excite the fundamental mechanical mode of a NW and
detect the resulting oscillations with the interferometer. Fig.~\ref{f:interf}(a)
shows the spectral response of the free-end displacement $x$ of the NW.
A main resonance and a smaller peak at lower frequency are clearly observable,
separated by 350~Hz. The asymmetric clamping of the NW to the Si chip,
realized by gluing the NW with one hexagonal facet in contact with the Si
surface (see Fig.~\ref{f:setup}(c)), splits the fundamental mode into a
doublet of flexural modes, oriented either perpendicular or parallel to the
Si surface. This interpretation is confirmed by a finite element model (FEM)
of the experimental system (see supporting information). The mode oscillating
perpendicular to the
surface is preferentially driven by the PZT, because its oscillation direction
coincides with the axis along which the PZT moves. This mode is also more
easily detected by the interferometer, since its direction of oscillation
coincides with the interferometer optical axis. For these reasons, we
interpret the main resonance in Fig.~\ref{f:interf}(a) as corresponding to
the perpendicular mode. The asymmetry visible in this resonance is due to the onset
of a weak mechanical non-linearity of the NW \cite{Nichol_nonlinear}.
When excited in the linear regime, each of these mechanical
resonances can be modeled as a driven, dissipative, harmonic
oscillator\cite{Montinaro}. Fitting the NW response using this model,
we extract for the perpendicular mode a resonant frequency
$\Omega_0 / 2\pi = 795.4$~kHz and a mechanical quality factor
$Q_{\perp} = 5800$ and for the parallel mode $Q_{\parallel} = 7600$.
Furthermore, by driving the main resonance as a function of the excitation
amplitude $V_{\text{PZT}}$, we explore the linear regime of the NW's free-end
displacement, as shown in Fig.~\ref{f:interf}(b). With a linear fit, we extract
a conversion factor, $\partial x / \partial V_{\text{PZT}} = 0.53 \pm 0.01$~nm/mV,
between the PZT drive amplitude and the amplitude of the free-end displacement.

\begin{figure}[tbhp]
	\includegraphics[width=3.33 in]{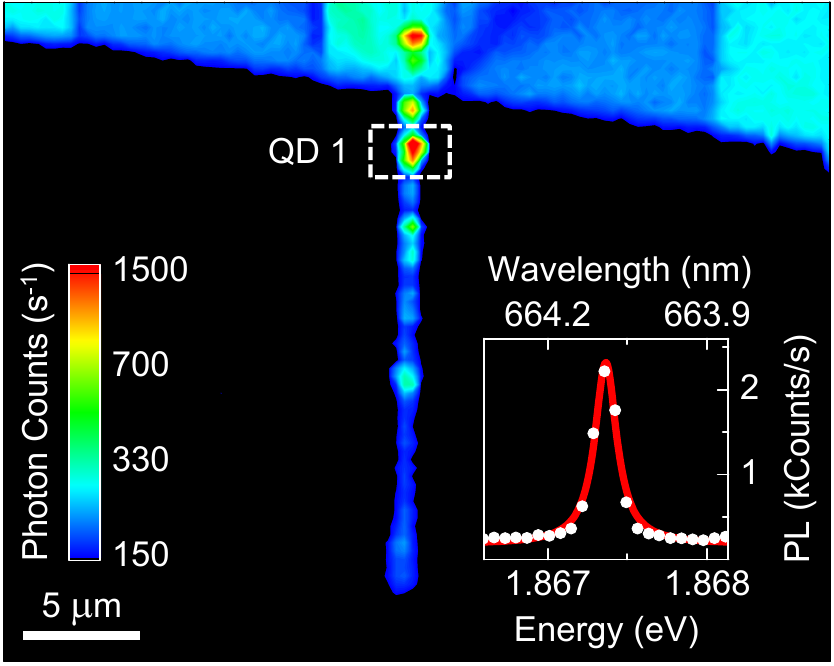}
	\caption{\label{f:PLmap} Spectrally filtered scanning confocal
	micrograph. As a function of the excitation laser position, we plot
	the light intensity detected from the sample (logarithmic color scale),
	spectrally filtered at the peak PL energy $E_{\text{ex}}^0 = 1.867$~eV,
	corresponding to exciton emission of QD 1. The inset shows the
	corresponding PL spectrum (white dots), together with a Lorentzian fit
	(red line). The linewidth (FWHM) is $140\,\mu$eV.}
\end{figure}

\begin{figure}[tbhp]
	\includegraphics[width=7 in]{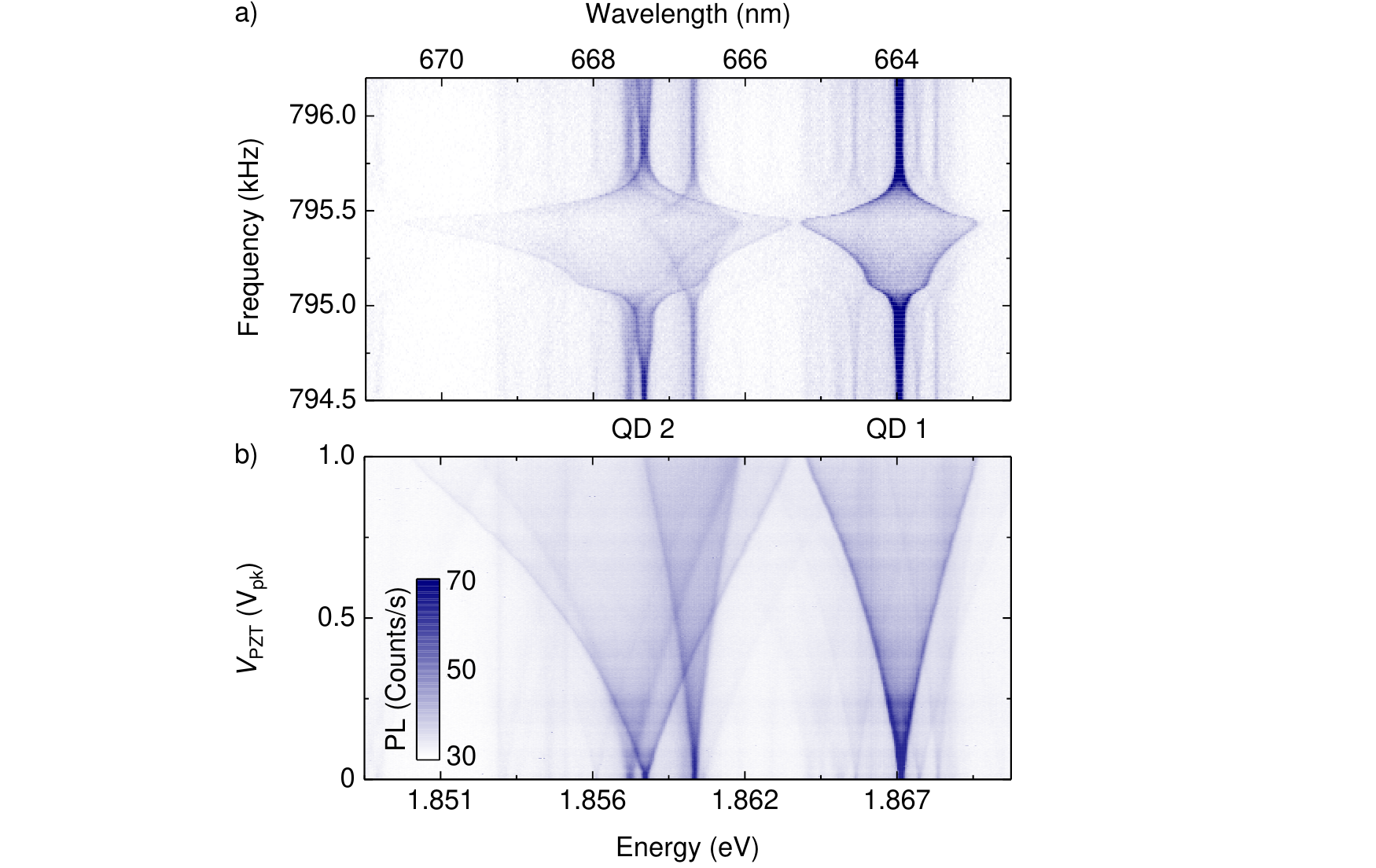}
	\caption{\label{f:PL} Effect of the NW excitation on the QD 
	photoluminescence. PL spectra of some neighboring QDs (labeled QD 1
	and QD 2) acquired while sweeping (a) the frequency of the PZT
	excitation, with $V_{\text{PZT}} = 1$~V$_{\text{pk}}$, and (b) the
	amplitude of the excitation, with the frequency set to the
	resonance of the NW's perpendicular flexural oscillation
	($\Omega = \Omega_0 = 2\pi \times 795.4$~kHz).}
\end{figure}

We study the opto-mechanical coupling by collecting PL from
individual QDs within a single NW. 
QDs in proximity of the clamped end of the NW have the
largest energy modulation, since the oscillation-induced material strain
is highest in this area (Fig.~\ref{f:setup}(b)). Using
the scanning confocal microscope, a number of suitable QDs are
identified near the clamped NW end, having bright, narrow, and
spectrally isolated exciton emission lines. Fig.~\ref{f:PLmap} shows a
spatial map of the PL at 1.867~eV (664~nm) under non-resonant laser
excitation of the sample. The plot also includes a weak component of
reflected light at the filtered energy, which reveals the position of
the NW and the Si substrate with its alignment markers. The map highlights
a conveniently located QD, which we label QD 1, whose PL spectral signature
includes an exciton emission peak, shown in the inset. In the next step,
the laser beam is maintained in alignment with QD 1's position and its PL spectrum
is recorded as a function of the PZT excitation frequency $\Omega / (2 \pi)$,
while holding the amplitude $V_{\text{PZT}}$ constant. As shown in
Fig.~\ref{f:PL}(a), several emission peaks are detected within the same
laser detection spot. As $\Omega$ is swept through the NW resonance
$\Omega_0$, the exciton emission peaks are broadened and deformed as a
consequence of the time-integrated sinusoidal motion of the NW\cite{Yeo}.
The envelope of the PL spectra as a function of $\Omega$ resembles the NW
displacement spectrum shown in Fig.~\ref{f:interf}(a). In particular, the
low-frequency shoulder of the broadened envelope corresponds to the
oscillation mode parallel to the Si surface.

We explore the range of the exciton energy modulation by
recording PL spectra as a function of the excitation amplitude
$V_{\text{PZT}}$, while driving the NW on resonance with the dominant
perpendicular mode ($\Omega = \Omega_0$).  As shown in
Fig.~\ref{f:PL}(b), each spectral line exhibits a different
broadening, as a consequence of its specific sensitivity to the local
strain. For high excitation voltages, we observe an asymmetric energy
broadening, due to the different response of the QD band structure
under compressive or tensile stress in the NW\cite{Bryant, Signorello}.
Note that a further increase of the excitation amplitude leads to a
saturation of the peak-to-peak exciton modulation width just beyond 14~meV.
It is currently not known whether this modulation is limited merely by how
hard we are able to drive the NW motion, or whether a more fundamental
saturation eventually limits the range.

While the mechanical motion of the NW in this experiment is best
described in classical terms, individual PL peaks from an embedded QD
can be approximated as resulting from a quantum two-level system with
an exciton transition energy $E_{\text{ex}}(x)$ between ground and
excited states $\ket{g}$ and $\ket{e}$\cite{Heiss}. The coupling between the NW
motion and the QD can then be introduced as a shift in the exciton
energy that depends on the displacement $x$ of the NW's free end.
Considering only the prominent perpendicular flexural vibration and
neglecting non-linear terms in $x$ (as in Ref.~\citenum{Treutlein}),
the time-dependent Hamiltonian of our hybrid system can be written as:
\begin{equation} 
	\hat{H}(t) = \frac{1}{2} m \dot{x}^2 + \frac{1}{2} m \Omega_0^2 x^2 + E_{\text{ex}}^0\,\frac{\hat{\sigma_z}}{2} + \left.\frac{\partial E_{\text{ex}}}{\partial x}\right|_{x=0} x\,\frac{\hat{\sigma_z}}{2},
	\label{e:H}
\end{equation}
where the first two terms describe the mechanical energy of the
unperturbed NW, the third term describes the emission energy of the
unperturbed QD, and the last one describes the opto-mechanical
interaction. In the equation, $m$ is the NW motional mass,
$E_{\text{ex}}^0$ is the transition energy of a QD exciton for the
NW at its rest position, $\hat{\sigma_z} = \ket{e}\bra{e} - \ket{g}\bra{g}$
is the Pauli operator of the two-level system, and
$\left. \frac{\partial E_{\text{ex}}}{\partial x} \right |_{x=0}$
is the opto-mechanical coupling parameter at the NW rest position. 
The NW motion produces a time-varying deformation of the NW's crystalline
structure, in turn altering the energy levels of the embedded QD via a
deformation potential, and resulting in a time-varying shift in the QD
exciton emission energies. The sign and magnitude of this shift under
compressive or tensile strain depend on the localization of the QD within the NW cross
section and possibly on intrinsic properties of each QD \cite{Jons}.



\begin{figure}[tbhp]
	\includegraphics[width=3.33 in]{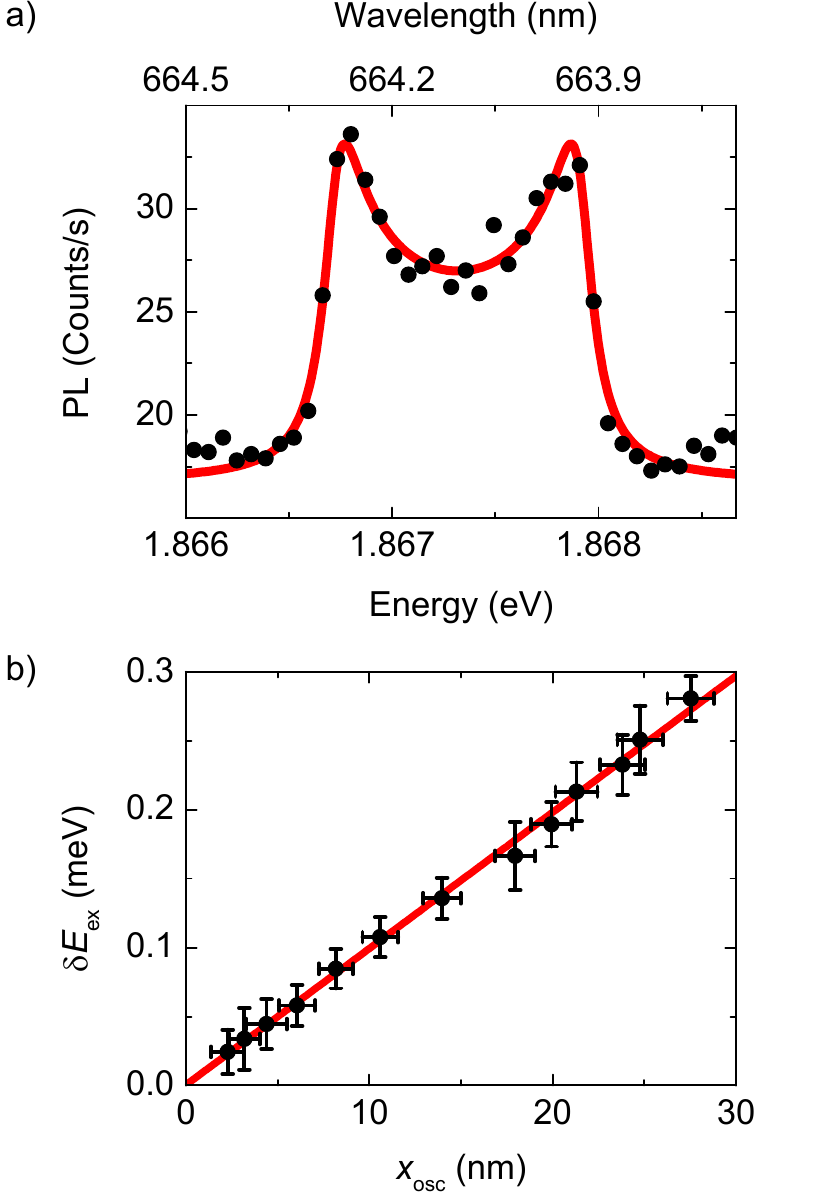}
	\caption{\label{f:g} Strength of the opto-mechanical coupling.
          (a)~PL spectrum of QD 1 (black dots) under NW excitation on
          resonance with the perpendicular flexural mode ($\Omega =
          \Omega_0$, $V_{\text{PZT}} = 250$~mV$_{\text{pk}}$). The red
          line is a model fit (see main text), from which the exciton
          energy shift amplitude $\delta E_{\text{ex}}$ is extracted.
          (b)~$\delta E_{\text{ex}}$ of QD 2 versus the NW free-end
          displacement amplitude $x_{\text{osc}}$. The red line is a
          linear fit, from which we extract the opto-mechanical
          coupling parameter $\left. \frac{\partial
              E_{\text{ex}}}{\partial x} \right |_{x=0} = 9.9 \pm
          0.7\,\mu$eV/nm.  The error bars on $x_{\text{osc}}$ are the same as
          mentioned in Fig.~\ref{f:interf}(b); those on $\delta
          E_{\text{ex}}$ are the standard deviations extracted from
          the fits of the mechanically excited PL spectra, as in (a).}
\end{figure}

To evaluate the strength of the opto-mechanical coupling, we extract
the PL profiles of the exciton lines for various values of the drive
$V_{\text{PZT}}$, e.g. Fig.~\ref{f:g}(a). The profiles are then fit
with a Lorentzian whose central energy $E_{\text{ex}}^0$ is modulated
by a sinusoid of amplitude $\delta E_{\text{ex}}$\cite{Arcizet}. Using
our interferometer measurements (Fig.~\ref{f:interf}(b)), we then
relate the displacement amplitude $x_{\text{osc}}$ of the NW free end
with the amplitude $\delta E_{\text{ex}}$. The result, displayed in
Fig.~\ref{f:g}(b) for QD 2 (which resides in the same optical spot as
QD 1), shows that in the linear regime of mechanical excitation,
$\delta E_{\text{ex}}$ is also linear in $x_{\text{osc}}$. A fit to
this data provides an opto-mechanical coupling parameter
$\left. \frac{\partial E_{\text{ex}}}{\partial x} \right |_{x=0} = 9.9
\pm 0.7\,\mu$eV/nm, which is one of the largest observed in our
measurements.

The energy shift of a QD exciton can be modeled by considering the
strain-dependent band structure of a semiconductor \cite{Chandra, van_de_walle}.
The deformation potentials and Poisson ratio have been recently measured
in an experiment on Zinc-Blende GaAs/AlGaAs core/shell NWs grown along
$\left\langle 1\,1\,1\right\rangle$\cite{Signorello}. These parameters
and a FEM of the NW strain tensor at the position of the QD in question
have been used to estimate the displacement-dependent energy shift. The
result of $13 \pm 2\,\mu$eV/nm is in agreement with our measurement and
corroborates the strain-dependence of the band structure as the dominant
coupling mechanism (see supporting information).

\begin{figure}[tbhp]
	\includegraphics[width=7 in]{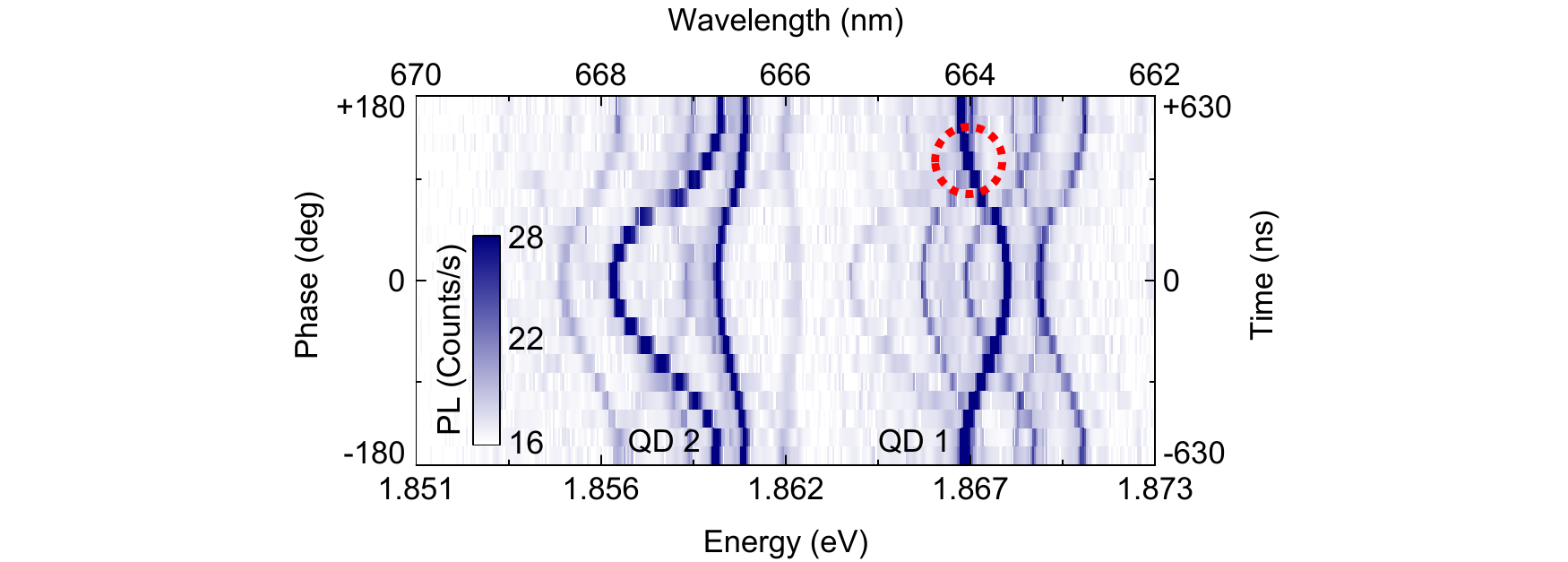}
	\caption{\label{f:strobo} Time-resolved PL evolution. Stroboscopic
	PL spectra of several neighboring QDs as a function of the phase
	(left axis) and the time delay (right axis) between the
	excitation-laser modulation and the PZT drive ($\Omega = \Omega_0$,
	$V_{\text{PZT}} = 250$~mV$_{\text{pk}}$). The dashed circle
	highlights two exciton spectral lines dynamically tuned to the same
	energy.}
\end{figure}

We study the time evolution of the QD exciton energy shift by
acquiring stroboscopic PL spectra. Two synchronized and isochronous
signals drive the NW on resonance through the PZT and, using an
acousto-optic modulator (AOM), chop the laser excitation with a
$5\%$~duty-cycle. The QDs are therefore excited only for $5\%$ of the
mechanical oscillation period of the NW. By recording PL spectra as a
function of the phase between the two modulation signals, as shown in
Fig.~\ref{f:strobo}, we explore the temporal evolution of the QD
exciton lines during a NW oscillation period. This experiment reveals
exciton lines, such as those of QD 1 and QD 2 in Fig.~\ref{f:strobo},
that respond to the mechanical oscillation of the NW with opposite
shifts in emission energy. The shifts in energy induced by strain are
a consequence of the change in the fundamental bands resulting from the
compression or extension of the lattice constant. Therefore, for a
given strain, exciton transitions from the same QD should show energy
shifts of the same sign and similar magnitude.  Conversely, emission
lines showing drastically different shift amplitudes or even shifts
with different signs correspond to QDs located at different positions
within the NW cross-section. In particular, two identical QDs within the same
optical collection spot, located on opposite sides of the NW neutral
axis, result in opposing strains produced for the same cantilever
free-end displacement. On the other hand, differences in the extension
and composition of each QD may also account for the varying responses
to NW motion \cite{Jons}. In either case, when two spatially and
spectrally close QD excitons display strong opto-mechanical couplings
of opposite sign, their energies may become degenerate for a
particular time in the oscillation period (or equivalently for a
particular position of the NW free end), as for the spectral lines
outlined by the dashed circle in Fig.~\ref{f:strobo}.  In the future,
exploiting this mechanically mediated tuning may allow us to couple
two nearby QDs within a single NW. In addition, the sinusoidal time evolution of the
PL spectral lines emerging from the measurement provides a
confirmation of the mechanical origin of the QD emission
broadening. Note that the modulation of the QD energy has no
measurable influence on the corresponding PL intensity.

In order to compare our results with other hybrid quantum
systems\cite{Treutlein, Yeo}, the opto-me\-cha\-ni\-cal interaction
described in Eq.~\ref{e:H} can also be expressed in terms of the
coupling rate
\begin{equation}
	\lambda \equiv \frac{1}{2\hbar} \left.\frac{\partial E_{\text{ex}}}{\partial x}\right|_{x=0} x_{\text{\tiny{ZPF}}},
	\label{e:lambda}
\end{equation}
which is the exciton frequency shift per vibrational quantum. Here,
$x_{\text{\tiny{ZPF}}} = \sqrt{\frac{\hbar}{2m\Omega_0}}$ is the NW's
zero-point motion at its free end and $\hbar$ is Planck's
constant. Using the FEM of the NW, we calculate its motional mass $m$,
which -- combined with knowledge of $\Omega_0$ -- allows us to
calculate $x_{\text{\tiny{ZPF}}} = (5.5 \pm 0.6)\times
10^{-14}$~m. Therefore, for QD 2, the coupling rate $\lambda / 2\pi =
66 \pm 12$~kHz. This opto-mechanical coupling rate is similar to that
recently measured by Yeo \textit{et al.}\cite{Yeo} for etched
nano-pillars containing self-assembled QDs, where $\lambda / 2\pi =
230 \pm 50$~kHz (note that in Ref.~\citenum{Yeo} $g_0 = 2 \lambda$).

Both here and in Yeo \textit{et al.}, the ratio
$\lambda / \Omega_0$ is not far from unity, which makes these kinds of
systems particularly promising for the quantum non-demolition (QND)
readout of a QD state through a precise measurement of the NW
displacement \cite{Yeo}.  In particular, using Eqs.~\ref{e:H} and
\ref{e:lambda}, we find that the displacement between the rest positions
of the NW free end in the QD states $\ket{g}$ and $\ket{e}$ is $4
x_{\text{\tiny{ZPF}}} \lambda / \Omega_0$.  This displacement, in
order to be observable, must be larger than $x_{\text{\tiny{ZPF}}}$;
in fact, at a finite temperature $T$, the displacement must be larger
than the NW's thermal fluctuations $x_{\text{\tiny{th}}}$.  This
implies that a determination of the QD state can be made through a
displacement measurement, if $\lambda / \Omega_0 > \frac{1}{4} \sqrt{1
  + 2 N}$, where $N$ is the average phonon occupation number of the
NW's fundamental mode.  In the high temperature limit $k_B T \gg \hbar
\Omega_0$, the requirement is that $\lambda / \Omega_0 >
\sqrt{\frac{k_B T}{8 \hbar \Omega_0}}$, where $k_B$ is the Boltzmann
constant. However, for our experimental parameters, the ratio
$\lambda / \Omega_0$ is still $10^3$ times too small for such effects
to be observed.

Auffeves and Richard \cite{auffeves} have recently proposed an alternative
approach to such a non-de\-mo\-li\-tion measurement, which takes
advantage of the high $Q$ of the NW oscillator. In their scheme, the
QD is optically excited by a continuous-wave laser modulated at the
NW resonance frequency. This process builds up, through constructive
interference, a large coherent mechanical excitation of the NW.
On resonance with a QD transition, the amplitude of the excitation is
roughly $Q$ times larger than the displacement difference calculated
in the aforementioned static case. For our experimental parameters,
this amplitude would be 6 times larger than the NW thermal fluctuations,
making it detectable by a high-sensitivity interferometer\cite{Nichol}.
It should be noted that a QND measurement also requires the time necessary to
build up such a coherent phonon field ($T_r$) to be smaller than the
quantum transition lifetime ($\tau_{\text{\tiny{ex}}}$), which is not
the case here ($T_r \approx 18$~ms, while $\tau_{\text{\tiny{ex}}}
\approx 1$~ns) nor in the experiment of Yeo \textit{et al.}\cite{Yeo}. The use
of a longer-lived QD state such as a dark exciton ($1\,\mu$s
\cite{McFarlane}) or a spin state ($0.5$~s \cite{Bar-Gill})
could bring the system closer to the required lifetime. In addition, given a
detection of the NW displacement with a large enough signal-to-noise
ratio, $T_r$ could be reduced using feedback damping, which
can modify a mechanical oscillator's response time without affecting
its intrinsic properties\cite{Poggio:2007}.

We note that prospects of quantum control over a mechanical resonator,
or proposals for using a mechanical resonator as a transducer for
quantum information, require the hybrid interaction to be large
compared to the rates at which the coupled systems decohere into their
local environments \cite{Groblacher}. Some proposals require the
condition of ``large cooperativity''\cite{clerk, hammerer}: $\lambda /
\sqrt{\gamma_{\text{\tiny{ex}}}\Gamma_{\text{\tiny{th}}}} > 1$, where
$\gamma_{\text{\tiny{ex}}}$ is the decoherence rate of the quantum
transition, in our case associated to a QD exciton ($>
1$~GHz\cite{Heiss}) and $\Gamma_{\text{\tiny{th}}} = \frac{k_B
  T}{\hbar Q}$ is the mechanical heating rate. The cooperativity in
this experiment is $10^{-3}$. Nevertheless, the QD-in-NW system is
particularly promising given that $\lambda$ could be improved by a
factor 2 (or bigger) by driving the NW at its second order (or higher)
flexural mode (see supporting information). Assuming that the
experiment can be carried out in a dilution refrigerator at $T =
10$~mK and that the mechanical $Q$ can be improved to a few times
$10^6$ -- perhaps by surface treatment, as was demonstrated in Si
cantilevers with similar aspect ratios\cite{rast}-- the large
cooperativity limit would then become accessible.

In summary, we demonstrate an ``as-grown'' opto-mechanical system
produced entirely by bottom-up self-assembly.  The structure's
intrinsic properties couple multiple QDs to the same NW mechanical
oscillator.  This interaction enables the tuning of QD energies over a
broad range exceeing 14 meV, opening the way for mechanically induced
coupling between different QDs in the NW. The sensitivity of the QDs
in our system to the resonant vibration of the NW could also be used
to reveal variations in the mechanical resonance frequency due to the
application of electrical or magnetic forces or to a change of the
mass of the NW. This fact opens the perspective of using our QD-in-NW
system as an integrated force probe or as a nanomechanical mass
sensor. By measuring the QD PL, one could monitor the NW motion in a
technically simpler way than optical interferometry \cite{Nichol,
  Carr, Belov, Favero} or other schemes\cite{Treacy, Montague, Hoch}.
 

\subsection{Supporting Information Available}
The mechanical properties of the NW, the effect of strain on the
QD-in-NW exciton energy, and the interferometer calibration are
discussed in detail. This material is available free of charge via the
Internet at
\href{http://poggiolab.unibas.ch/full/NWQDSuppInfo.pdf}{http://poggiolab.unibas.ch/full/NWQDSuppInfo.pdf}.


\subsection{Author Contributions}
M.Mo., Y.F., A.F.M., and M.P. conceived the experiment. E.R.A., M.H.,
and A.F.M. synthesized the NWs. G.W. and R.W. designed and set up the
confocal scanning microscope. M.Mo., G.W., and M.Mu. performed the
measurements, under the supervision of M.P. and R.W.. M.Mo. analyzed
the data and performed the FEM simulations.  M.Mo and M.P. wrote the
manuscript. All authors discussed the results and contributed to the
manuscript.


\subsection{Acknowledgements}
The authors thank Dr. Jean Teissier and Dr. Fei Xue for fruitful
discussion and Dr. Pengfei Wang, Benedikt E. Herzog, Dennis P. Weber,
Andrea Mehlin, and Davide Cadeddu for technical support. We acknowledge
support from the Canton Aargau, the Swiss NSF (Grant No. 200020-140478),
the National Center of Competence in Research for Quantum Science and
Technology (NCCR-QSIT), the D-A-CH program of the Swiss NSF (Grant No. 132506),
the ERC Starting Grant UpCon, and the ERC Starting Grant NWScan (Grant
No. 334762).



\bibliography{bibfile}

\providecommand*\mcitethebibliography{\thebibliography}
\csname @ifundefined\endcsname{endmcitethebibliography}
  {\let\endmcitethebibliography\endthebibliography}{}
\begin{mcitethebibliography}{45}
\providecommand*\natexlab[1]{#1}
\providecommand*\mciteSetBstSublistMode[1]{}
\providecommand*\mciteSetBstMaxWidthForm[2]{}
\providecommand*\mciteBstWouldAddEndPuncttrue
  {\def\EndOfBibitem{\unskip.}}
\providecommand*\mciteBstWouldAddEndPunctfalse
  {\let\EndOfBibitem\relax}
\providecommand*\mciteSetBstMidEndSepPunct[3]{}
\providecommand*\mciteSetBstSublistLabelBeginEnd[3]{}
\providecommand*\EndOfBibitem{}
\mciteSetBstSublistMode{f}
\mciteSetBstMaxWidthForm{subitem}{(\alph{mcitesubitemcount})}
\mciteSetBstSublistLabelBeginEnd
  {\mcitemaxwidthsubitemform\space}
  {\relax}
  {\relax}

\bibitem[Teufel et~al.(2011)Teufel, Donner, Li, Harlow, Allman, Cicak, Sirois,
  Whittaker, Lehnert, and Simmonds]{Teufel}
Teufel,~J.~D.; Donner,~T.; Li,~D.; Harlow,~J.~W.; Allman,~M.~S.; Cicak,~K.;
  Sirois,~A.~J.; Whittaker,~J.~D.; Lehnert,~K.~W.; Simmonds,~R.~W.
  \emph{Nature} \textbf{2011}, \emph{475}, 359--363\relax
\mciteBstWouldAddEndPuncttrue
\mciteSetBstMidEndSepPunct{\mcitedefaultmidpunct}
{\mcitedefaultendpunct}{\mcitedefaultseppunct}\relax
\EndOfBibitem
\bibitem[Chan et~al.(2011)Chan, Alegre, Safavi-Naeini, Hill, Krause,
  Gr\"oblacher, Aspelmeyer, and Painter]{Chan}
Chan,~J.; Alegre,~T. P.~M.; Safavi-Naeini,~A.~H.; Hill,~J.~T.; Krause,~A.;
  Gr\"oblacher,~S.; Aspelmeyer,~M.; Painter,~O. \emph{Nature} \textbf{2011},
  \emph{478}, 89--92\relax
\mciteBstWouldAddEndPuncttrue
\mciteSetBstMidEndSepPunct{\mcitedefaultmidpunct}
{\mcitedefaultendpunct}{\mcitedefaultseppunct}\relax
\EndOfBibitem
\bibitem[O'Connell et~al.(2010)O'Connell, Hofheinz, Ansmann, Bialczak,
  Lenander, Lucero, Neeley, Sank, Wang, Weides, Wenner, Martinis, and
  Cleland]{OConnell}
O'Connell,~A.~D.; Hofheinz,~M.; Ansmann,~M.; Bialczak,~R.~C.; Lenander,~M.;
  Lucero,~E.; Neeley,~M.; Sank,~D.; Wang,~H.; Weides,~M.; Wenner,~J.;
  Martinis,~J.~M.; Cleland,~A.~N. \emph{Nature} \textbf{2010}, \emph{464},
  697--703\relax
\mciteBstWouldAddEndPuncttrue
\mciteSetBstMidEndSepPunct{\mcitedefaultmidpunct}
{\mcitedefaultendpunct}{\mcitedefaultseppunct}\relax
\EndOfBibitem
\bibitem[Treutlein et~al.(2012)Treutlein, Genes, Hammerer, Poggio, and
  Rabl]{Treutlein}
Treutlein,~P.; Genes,~C.; Hammerer,~K.; Poggio,~M.; Rabl,~P.
  \emph{{arXiv:1210.4151}} \textbf{2012}, \relax
\mciteBstWouldAddEndPunctfalse
\mciteSetBstMidEndSepPunct{\mcitedefaultmidpunct}
{}{\mcitedefaultseppunct}\relax
\EndOfBibitem
\bibitem[Rabl et~al.(2010)Rabl, Kolkowitz, Koppens, Harris, Zoller, and
  Lukin]{Rabl}
Rabl,~P.; Kolkowitz,~S.~J.; Koppens,~F. H.~L.; Harris,~J. G.~E.; Zoller,~P.;
  Lukin,~M.~D. \emph{Nature Physics} \textbf{2010}, \emph{6}, 602--608\relax
\mciteBstWouldAddEndPuncttrue
\mciteSetBstMidEndSepPunct{\mcitedefaultmidpunct}
{\mcitedefaultendpunct}{\mcitedefaultseppunct}\relax
\EndOfBibitem
\bibitem[Kolkowitz et~al.(2012)Kolkowitz, Bleszynski~Jayich, Unterreithmeier,
  Bennett, Rabl, Harris, and Lukin]{Kolkowitz}
Kolkowitz,~S.; Bleszynski~Jayich,~A.~C.; Unterreithmeier,~Q.~P.;
  Bennett,~S.~D.; Rabl,~P.; Harris,~J. G.~E.; Lukin,~M.~D. \emph{Science}
  \textbf{2012}, \emph{335}, 1603--1606\relax
\mciteBstWouldAddEndPuncttrue
\mciteSetBstMidEndSepPunct{\mcitedefaultmidpunct}
{\mcitedefaultendpunct}{\mcitedefaultseppunct}\relax
\EndOfBibitem
\bibitem[{McGee} et~al.(2013){McGee}, Meiser, Regal, Lehnert, and
  Holland]{McGee}
{McGee},~S.~A.; Meiser,~D.; Regal,~C.~A.; Lehnert,~K.~W.; Holland,~M.~J.
  \emph{Physical Review A} \textbf{2013}, \emph{87}, 053818\relax
\mciteBstWouldAddEndPuncttrue
\mciteSetBstMidEndSepPunct{\mcitedefaultmidpunct}
{\mcitedefaultendpunct}{\mcitedefaultseppunct}\relax
\EndOfBibitem
\bibitem[Palomaki et~al.(2013)Palomaki, Harlow, Teufel, Simmonds, and
  Lehnert]{Palomaki}
Palomaki,~T.~A.; Harlow,~J.~W.; Teufel,~J.~D.; Simmonds,~R.~W.; Lehnert,~K.~W.
  \emph{Nature} \textbf{2013}, \emph{495}, 210--214\relax
\mciteBstWouldAddEndPuncttrue
\mciteSetBstMidEndSepPunct{\mcitedefaultmidpunct}
{\mcitedefaultendpunct}{\mcitedefaultseppunct}\relax
\EndOfBibitem
\bibitem[Anetsberger et~al.(2010)Anetsberger, Gavartin, Arcizet,
  Unterreithmeier, Weig, Gorodetsky, Kotthaus, and Kippenberg]{Anetsberger}
Anetsberger,~G.; Gavartin,~E.; Arcizet,~O.; Unterreithmeier,~Q.~P.;
  Weig,~E.~M.; Gorodetsky,~M.~L.; Kotthaus,~J.~P.; Kippenberg,~T.~J.
  \emph{Physical Review A} \textbf{2010}, \emph{82}, 061804\relax
\mciteBstWouldAddEndPuncttrue
\mciteSetBstMidEndSepPunct{\mcitedefaultmidpunct}
{\mcitedefaultendpunct}{\mcitedefaultseppunct}\relax
\EndOfBibitem
\bibitem[Bochmann et~al.(2013)Bochmann, Vainsencher, Awschalom, and
  Cleland]{Bochmann}
Bochmann,~J.; Vainsencher,~A.; Awschalom,~D.~D.; Cleland,~A.~N. \emph{Nature
  Physics} \textbf{2013}, \emph{9}, 712--716\relax
\mciteBstWouldAddEndPuncttrue
\mciteSetBstMidEndSepPunct{\mcitedefaultmidpunct}
{\mcitedefaultendpunct}{\mcitedefaultseppunct}\relax
\EndOfBibitem
\bibitem[Armour et~al.(2002)Armour, Blencowe, and Schwab]{Armour}
Armour,~A.~D.; Blencowe,~M.~P.; Schwab,~K.~C. \emph{Physical Review Letters}
  \textbf{2002}, \emph{88}, 148301\relax
\mciteBstWouldAddEndPuncttrue
\mciteSetBstMidEndSepPunct{\mcitedefaultmidpunct}
{\mcitedefaultendpunct}{\mcitedefaultseppunct}\relax
\EndOfBibitem
\bibitem[Camerer et~al.(2011)Camerer, Korppi, J\"ockel, Hunger, H\"ansch, and
  Treutlein]{Camerer}
Camerer,~S.; Korppi,~M.; J\"ockel,~A.; Hunger,~D.; H\"ansch,~T.~W.;
  Treutlein,~P. \emph{Physical Review Letters} \textbf{2011}, \emph{107},
  223001\relax
\mciteBstWouldAddEndPuncttrue
\mciteSetBstMidEndSepPunct{\mcitedefaultmidpunct}
{\mcitedefaultendpunct}{\mcitedefaultseppunct}\relax
\EndOfBibitem
\bibitem[Yeo et~al.(2014)Yeo, de~Assis, Gloppe, Dupont-Ferrier, Verlot, Malik,
  Dupuy, Claudon, G\'erard, Auff\`{e}ves, Nogues, Seidelin, Poizat, Arcizet,
  and Richard]{Yeo}
Yeo,~I.; de~Assis,~P.-L.; Gloppe,~A.; Dupont-Ferrier,~E.; Verlot,~P.;
  Malik,~N.~S.; Dupuy,~E.; Claudon,~J.; G\'erard,~J.-M.; Auff\`{e}ves,~A.;
  Nogues,~G.; Seidelin,~S.; Poizat,~J.-P.; Arcizet,~O.; Richard,~M.
  \emph{Nature Nanotechnology} \textbf{2014}, \emph{9}, 106--110\relax
\mciteBstWouldAddEndPuncttrue
\mciteSetBstMidEndSepPunct{\mcitedefaultmidpunct}
{\mcitedefaultendpunct}{\mcitedefaultseppunct}\relax
\EndOfBibitem
\bibitem[Arcizet et~al.(2011)Arcizet, Jacques, Siria, Poncharal, Vincent, and
  Seidelin]{Arcizet}
Arcizet,~O.; Jacques,~V.; Siria,~A.; Poncharal,~P.; Vincent,~P.; Seidelin,~S.
  \emph{Nature Physics} \textbf{2011}, \emph{7}, 879--883\relax
\mciteBstWouldAddEndPuncttrue
\mciteSetBstMidEndSepPunct{\mcitedefaultmidpunct}
{\mcitedefaultendpunct}{\mcitedefaultseppunct}\relax
\EndOfBibitem
\bibitem[Teissier et~al.(2014)Teissier, Barfuss, Appel, Neu, and
  Maletinsky]{Teissier}
Teissier,~J.; Barfuss,~A.; Appel,~P.; Neu,~E.; Maletinsky,~P.
  \emph{{arXiv:1403.3405}} \textbf{2014}, \relax
\mciteBstWouldAddEndPunctfalse
\mciteSetBstMidEndSepPunct{\mcitedefaultmidpunct}
{}{\mcitedefaultseppunct}\relax
\EndOfBibitem
\bibitem[Ovartchaiyapong et~al.(2014)Ovartchaiyapong, Lee, Myers, and
  Bles\-zyn\-ski Jayich]{Jayich}
Ovartchaiyapong,~P.; Lee,~K.~W.; Myers,~B.~A.; Bles\-zyn\-ski Jayich,~A.~C.
  \emph{{arXiv:1403.4173}} \textbf{2014}, \relax
\mciteBstWouldAddEndPunctfalse
\mciteSetBstMidEndSepPunct{\mcitedefaultmidpunct}
{}{\mcitedefaultseppunct}\relax
\EndOfBibitem
\bibitem[Heiss et~al.(2013)Heiss, Fontana, Gustafsson, W\"ust, Magen, O'Regan,
  Luo, Ketterer, Conesa-Boj, Kuhlmann, Houel, Russo-Averchi, Morante, Cantoni,
  Marzari, Arbiol, Zunger, Warburton, and Fontcuberta~i Morral]{Heiss}
Heiss,~M. et~al.  \emph{Nature Materials} \textbf{2013}, \emph{12},
  439--444\relax
\mciteBstWouldAddEndPuncttrue
\mciteSetBstMidEndSepPunct{\mcitedefaultmidpunct}
{\mcitedefaultendpunct}{\mcitedefaultseppunct}\relax
\EndOfBibitem
\bibitem[Uccelli et~al.(2011)Uccelli, Arbiol, Magen, Krogstrup, Russo-Averchi,
  Heiss, Mugny, Morier-Genoud, Nyg{\aa}rd, Morante, and Fontcuberta~i
  Morral]{uccelli}
Uccelli,~E.; Arbiol,~J.; Magen,~C.; Krogstrup,~P.; Russo-Averchi,~E.;
  Heiss,~M.; Mugny,~G.; Morier-Genoud,~F.; Nyg{\aa}rd,~J.; Morante,~J.~R.;
  Fontcuberta~i Morral,~A. \emph{Nano Letters} \textbf{2011}, \emph{11},
  3827--3832\relax
\mciteBstWouldAddEndPuncttrue
\mciteSetBstMidEndSepPunct{\mcitedefaultmidpunct}
{\mcitedefaultendpunct}{\mcitedefaultseppunct}\relax
\EndOfBibitem
\bibitem[Russo-Averchi et~al.(2012)Russo-Averchi, Heiss, Michelet, Krogstrup,
  Nyg{\aa}rd, Magen, Morante, Uccelli, Arbiol, and Fontcuberta~i Morral]{Russo}
Russo-Averchi,~E.; Heiss,~M.; Michelet,~L.; Krogstrup,~P.; Nyg{\aa}rd,~J.;
  Magen,~C.; Morante,~J.~R.; Uccelli,~E.; Arbiol,~J.; Fontcuberta~i Morral,~A.
  \emph{Nanoscale} \textbf{2012}, \emph{4}, 1486--1490\relax
\mciteBstWouldAddEndPuncttrue
\mciteSetBstMidEndSepPunct{\mcitedefaultmidpunct}
{\mcitedefaultendpunct}{\mcitedefaultseppunct}\relax
\EndOfBibitem
\bibitem[Heigoldt et~al.(2009)Heigoldt, Arbiol, Spirkoska, Rebled, Conesa-Boj,
  Abstreiter, Peir\'o, Morante, and Fontcuberta~i Morral]{heigoldt}
Heigoldt,~M.; Arbiol,~J.; Spirkoska,~D.; Rebled,~J.~M.; Conesa-Boj,~S.;
  Abstreiter,~G.; Peir\'o,~F.; Morante,~J.~R.; Fontcuberta~i Morral,~A.
  \emph{Journal of Materials Chemistry} \textbf{2009}, \emph{19}, 840\relax
\mciteBstWouldAddEndPuncttrue
\mciteSetBstMidEndSepPunct{\mcitedefaultmidpunct}
{\mcitedefaultendpunct}{\mcitedefaultseppunct}\relax
\EndOfBibitem
\bibitem[H\"ogele et~al.(2008)H\"ogele, Seidl, Kroner, Karrai, Schulhauser,
  Sqalli, Scrimgeour, and Warburton]{Hoegele}
H\"ogele,~A.; Seidl,~S.; Kroner,~M.; Karrai,~K.; Schulhauser,~C.; Sqalli,~O.;
  Scrimgeour,~J.; Warburton,~R.~J. \emph{Review of Scientific Instruments}
  \textbf{2008}, \emph{79}, 023709\relax
\mciteBstWouldAddEndPuncttrue
\mciteSetBstMidEndSepPunct{\mcitedefaultmidpunct}
{\mcitedefaultendpunct}{\mcitedefaultseppunct}\relax
\EndOfBibitem
\bibitem[Bruland et~al.(1999)Bruland, Garbini, Dougherty, Chao, Jensen, and
  Sidles]{Bruland}
Bruland,~K.~J.; Garbini,~J.~L.; Dougherty,~W.~M.; Chao,~S.~H.; Jensen,~S.~E.;
  Sidles,~J.~A. \emph{Review of Scientific Instruments} \textbf{1999},
  \emph{70}, 3542--3544\relax
\mciteBstWouldAddEndPuncttrue
\mciteSetBstMidEndSepPunct{\mcitedefaultmidpunct}
{\mcitedefaultendpunct}{\mcitedefaultseppunct}\relax
\EndOfBibitem
\bibitem[Nichol et~al.(2009)Nichol, Hemesath, Lauhon, and
  Budakian]{Nichol_nonlinear}
Nichol,~J.~M.; Hemesath,~E.~R.; Lauhon,~L.~J.; Budakian,~R. \emph{Applied
  Physics Letters} \textbf{2009}, \emph{95}, 123116\relax
\mciteBstWouldAddEndPuncttrue
\mciteSetBstMidEndSepPunct{\mcitedefaultmidpunct}
{\mcitedefaultendpunct}{\mcitedefaultseppunct}\relax
\EndOfBibitem
\bibitem[Montinaro et~al.(2012)Montinaro, Mehlin, Solanki, Peddibhotla, Mack,
  Awschalom, and Poggio]{Montinaro}
Montinaro,~M.; Mehlin,~A.; Solanki,~H.~S.; Peddibhotla,~P.; Mack,~S.;
  Awschalom,~D.~D.; Poggio,~M. \emph{Applied Physics Letters} \textbf{2012},
  \emph{101}, 133104\relax
\mciteBstWouldAddEndPuncttrue
\mciteSetBstMidEndSepPunct{\mcitedefaultmidpunct}
{\mcitedefaultendpunct}{\mcitedefaultseppunct}\relax
\EndOfBibitem
\bibitem[Bryant et~al.(2011)Bryant, Zieli\'nski, Malkova, Sims, Jask\'olski,
  and Aizpurua]{Bryant}
Bryant,~G.~W.; Zieli\'nski,~M.; Malkova,~N.; Sims,~J.; Jask\'olski,~W.;
  Aizpurua,~J. \emph{Physical Review B} \textbf{2011}, \emph{84}, 235412\relax
\mciteBstWouldAddEndPuncttrue
\mciteSetBstMidEndSepPunct{\mcitedefaultmidpunct}
{\mcitedefaultendpunct}{\mcitedefaultseppunct}\relax
\EndOfBibitem
\bibitem[Signorello et~al.(2013)Signorello, Karg, Bj\"ork, Gotsmann, and
  Riel]{Signorello}
Signorello,~G.; Karg,~S.; Bj\"ork,~M.~T.; Gotsmann,~B.; Riel,~H. \emph{Nano
  Letters} \textbf{2013}, \emph{13}, 917--924\relax
\mciteBstWouldAddEndPuncttrue
\mciteSetBstMidEndSepPunct{\mcitedefaultmidpunct}
{\mcitedefaultendpunct}{\mcitedefaultseppunct}\relax
\EndOfBibitem
\bibitem[J\"ons et~al.(2011)J\"ons, Hafenbrak, Singh, Ding, Plumhof, Rastelli,
  Schmidt, Bester, and Michler]{Jons}
J\"ons,~K.~D.; Hafenbrak,~R.; Singh,~R.; Ding,~F.; Plumhof,~J.~D.;
  Rastelli,~A.; Schmidt,~O.~G.; Bester,~G.; Michler,~P. \emph{Physical Review
  Letters} \textbf{2011}, \emph{107}, 217402\relax
\mciteBstWouldAddEndPuncttrue
\mciteSetBstMidEndSepPunct{\mcitedefaultmidpunct}
{\mcitedefaultendpunct}{\mcitedefaultseppunct}\relax
\EndOfBibitem
\bibitem[Chandrasekhar and Pollak(1977)Chandrasekhar, and Pollak]{Chandra}
Chandrasekhar,~M.; Pollak,~F.~H. \emph{Physical Review B} \textbf{1977},
  \emph{15}, 2127--2144\relax
\mciteBstWouldAddEndPuncttrue
\mciteSetBstMidEndSepPunct{\mcitedefaultmidpunct}
{\mcitedefaultendpunct}{\mcitedefaultseppunct}\relax
\EndOfBibitem
\bibitem[Van~de Walle(1989)]{van_de_walle}
Van~de Walle,~C.~G. \emph{Physical Review B} \textbf{1989}, \emph{39},
  1871--1883\relax
\mciteBstWouldAddEndPuncttrue
\mciteSetBstMidEndSepPunct{\mcitedefaultmidpunct}
{\mcitedefaultendpunct}{\mcitedefaultseppunct}\relax
\EndOfBibitem
\bibitem[Auff\`{e}ves and Richard(2013)Auff\`{e}ves, and Richard]{auffeves}
Auff\`{e}ves,~A.; Richard,~M. \emph{{arXiv:1305.4252}} \textbf{2013}, \relax
\mciteBstWouldAddEndPunctfalse
\mciteSetBstMidEndSepPunct{\mcitedefaultmidpunct}
{}{\mcitedefaultseppunct}\relax
\EndOfBibitem
\bibitem[Nichol et~al.(2008)Nichol, Hemesath, Lauhon, and Budakian]{Nichol}
Nichol,~J.~M.; Hemesath,~E.~R.; Lauhon,~L.~J.; Budakian,~R. \emph{Applied
  Physics Letters} \textbf{2008}, \emph{93}, 193110\relax
\mciteBstWouldAddEndPuncttrue
\mciteSetBstMidEndSepPunct{\mcitedefaultmidpunct}
{\mcitedefaultendpunct}{\mcitedefaultseppunct}\relax
\EndOfBibitem
\bibitem[{McFarlane} et~al.(2009){McFarlane}, Dalgarno, Gerardot, Hadfield,
  Warburton, Karrai, Badolato, and Petroff]{McFarlane}
{McFarlane},~J.; Dalgarno,~P.~A.; Gerardot,~B.~D.; Hadfield,~R.~H.;
  Warburton,~R.~J.; Karrai,~K.; Badolato,~A.; Petroff,~P.~M. \emph{Applied
  Physics Letters} \textbf{2009}, \emph{94}, 093113\relax
\mciteBstWouldAddEndPuncttrue
\mciteSetBstMidEndSepPunct{\mcitedefaultmidpunct}
{\mcitedefaultendpunct}{\mcitedefaultseppunct}\relax
\EndOfBibitem
\bibitem[Bar-Gill et~al.(2013)Bar-Gill, Pham, Jarmola, Budker, and
  Walsworth]{Bar-Gill}
Bar-Gill,~N.; Pham,~L.~M.; Jarmola,~A.; Budker,~D.; Walsworth,~R.~L.
  \emph{Nature Communications} \textbf{2013}, \emph{4}, 1743\relax
\mciteBstWouldAddEndPuncttrue
\mciteSetBstMidEndSepPunct{\mcitedefaultmidpunct}
{\mcitedefaultendpunct}{\mcitedefaultseppunct}\relax
\EndOfBibitem
\bibitem[Poggio et~al.(2007)Poggio, Degen, Mamin, and Rugar]{Poggio:2007}
Poggio,~M.; Degen,~C.~L.; Mamin,~H.~J.; Rugar,~D. \emph{Physical Review
  Letters} \textbf{2007}, \emph{99}, 017201\relax
\mciteBstWouldAddEndPuncttrue
\mciteSetBstMidEndSepPunct{\mcitedefaultmidpunct}
{\mcitedefaultendpunct}{\mcitedefaultseppunct}\relax
\EndOfBibitem
\bibitem[Gr\"oblacher et~al.(2009)Gr\"oblacher, Hammerer, Vanner, and
  Aspelmeyer]{Groblacher}
Gr\"oblacher,~S.; Hammerer,~K.; Vanner,~M.~R.; Aspelmeyer,~M. \emph{Nature}
  \textbf{2009}, \emph{460}, 724--727\relax
\mciteBstWouldAddEndPuncttrue
\mciteSetBstMidEndSepPunct{\mcitedefaultmidpunct}
{\mcitedefaultendpunct}{\mcitedefaultseppunct}\relax
\EndOfBibitem
\bibitem[Clerk et~al.(2008)Clerk, Marquardt, and Jacobs]{clerk}
Clerk,~A.~A.; Marquardt,~F.; Jacobs,~K. \emph{New Journal of Physics}
  \textbf{2008}, \emph{10}, 095010\relax
\mciteBstWouldAddEndPuncttrue
\mciteSetBstMidEndSepPunct{\mcitedefaultmidpunct}
{\mcitedefaultendpunct}{\mcitedefaultseppunct}\relax
\EndOfBibitem
\bibitem[Hammerer et~al.(2009)Hammerer, Aspelmeyer, Polzik, and
  Zoller]{hammerer}
Hammerer,~K.; Aspelmeyer,~M.; Polzik,~E.~S.; Zoller,~P. \emph{Physical Review
  Letters} \textbf{2009}, \emph{102}, 020501\relax
\mciteBstWouldAddEndPuncttrue
\mciteSetBstMidEndSepPunct{\mcitedefaultmidpunct}
{\mcitedefaultendpunct}{\mcitedefaultseppunct}\relax
\EndOfBibitem
\bibitem[Rast et~al.(2006)Rast, Gysin, Ruff, Gerber, Meyer, and Lee]{rast}
Rast,~S.; Gysin,~U.; Ruff,~P.; Gerber,~C.; Meyer,~E.; Lee,~D.~W.
  \emph{Nanotechnology} \textbf{2006}, \emph{17}, S189\relax
\mciteBstWouldAddEndPuncttrue
\mciteSetBstMidEndSepPunct{\mcitedefaultmidpunct}
{\mcitedefaultendpunct}{\mcitedefaultseppunct}\relax
\EndOfBibitem
\bibitem[Carr and Craighead(1997)Carr, and Craighead]{Carr}
Carr,~D.~W.; Craighead,~H.~G. \emph{Journal of Vacuum Science \& Technology B}
  \textbf{1997}, \emph{15}, 2760--2763\relax
\mciteBstWouldAddEndPuncttrue
\mciteSetBstMidEndSepPunct{\mcitedefaultmidpunct}
{\mcitedefaultendpunct}{\mcitedefaultseppunct}\relax
\EndOfBibitem
\bibitem[Belov et~al.(2008)Belov, Quitoriano, Sharma, Hiebert, Kamins, and
  Evoy]{Belov}
Belov,~M.; Quitoriano,~N.~J.; Sharma,~S.; Hiebert,~W.~K.; Kamins,~T.~I.;
  Evoy,~S. \emph{Journal of Applied Physics} \textbf{2008}, \emph{103},
  074304\relax
\mciteBstWouldAddEndPuncttrue
\mciteSetBstMidEndSepPunct{\mcitedefaultmidpunct}
{\mcitedefaultendpunct}{\mcitedefaultseppunct}\relax
\EndOfBibitem
\bibitem[Favero et~al.(2009)Favero, Stapfner, Hunger, Paulitschke, Reichel,
  Lorenz, Weig, and Karrai]{Favero}
Favero,~I.; Stapfner,~S.; Hunger,~D.; Paulitschke,~P.; Reichel,~J.; Lorenz,~H.;
  Weig,~E.~M.; Karrai,~K. \emph{Optics Express} \textbf{2009}, \emph{17},
  12813--12820\relax
\mciteBstWouldAddEndPuncttrue
\mciteSetBstMidEndSepPunct{\mcitedefaultmidpunct}
{\mcitedefaultendpunct}{\mcitedefaultseppunct}\relax
\EndOfBibitem
\bibitem[Treacy et~al.(1996)Treacy, Ebbesen, and Gibson]{Treacy}
Treacy,~M. M.~J.; Ebbesen,~T.~W.; Gibson,~J.~M. \emph{Nature} \textbf{1996},
  \emph{381}, 678--680\relax
\mciteBstWouldAddEndPuncttrue
\mciteSetBstMidEndSepPunct{\mcitedefaultmidpunct}
{\mcitedefaultendpunct}{\mcitedefaultseppunct}\relax
\EndOfBibitem
\bibitem[Montague et~al.(2011)Montague, Dalberth, Gray, Seghete, Bertness,
  George, Bright, Rogers, and Sanford]{Montague}
Montague,~J.~R.; Dalberth,~M.; Gray,~J.~M.; Seghete,~D.; Bertness,~K.~A.;
  George,~S.~M.; Bright,~V.~M.; Rogers,~C.~T.; Sanford,~N.~A. \emph{Sensors and
  Actuators A} \textbf{2011}, \emph{165}, 59--65\relax
\mciteBstWouldAddEndPuncttrue
\mciteSetBstMidEndSepPunct{\mcitedefaultmidpunct}
{\mcitedefaultendpunct}{\mcitedefaultseppunct}\relax
\EndOfBibitem
\bibitem[Hoch et~al.(2011)Hoch, Montague, Bright, Rogers, Bertness, Teufel, and
  Lehnert]{Hoch}
Hoch,~S.~W.; Montague,~J.~R.; Bright,~V.~M.; Rogers,~C.~T.; Bertness,~K.~A.;
  Teufel,~J.~D.; Lehnert,~K.~W. \emph{Applied Physics Letters} \textbf{2011},
  \emph{99}, 053101\relax
\mciteBstWouldAddEndPuncttrue
\mciteSetBstMidEndSepPunct{\mcitedefaultmidpunct}
{\mcitedefaultendpunct}{\mcitedefaultseppunct}\relax
\EndOfBibitem
\end{mcitethebibliography}

	



\end{document}